\documentclass[pdflatex,sn-mathphys-num]{sn-jnl}% Math and Physical Sciences Numbered Reference Style
\usepackage{graphicx}%
\usepackage{multirow}%
\usepackage{siunitx}
\usepackage{amsmath,amssymb,amsfonts}%
\usepackage{amsthm}%
\usepackage{mathrsfs}%
\usepackage[title]{appendix}%
\usepackage{xcolor}%
\usepackage{textcomp}%
\usepackage{manyfoot}%
\usepackage{booktabs}%
\usepackage{algorithm}%
\usepackage{algorithmicx}%
\usepackage{algpseudocode}%
\usepackage{tabularx}  % 必须添加的包
\usepackage{listings}%
\theoremstyle{thmstyleone}%
\theoremstyle{thmstyletwo}%

\theoremstyle{thmstylethree}%

\begin{document}

\title[Article Title]{Causal Graph Neural Networks for Mining Stable Disease Biomarkers}


\author*[1]{\fnm{Chaowang} \sur{Lan}}\email{chaowanglan@guet.edu.cn}
\equalcont{These authors contributed equally to this work.}

\author[1]{\fnm{Jingxin} \sur{Wu}}\email{1901610218@mails.guet.edu.cn}
\equalcont{These authors contributed equally to this work.}

\author[1]{\fnm{Yulong} \sur{Yuan}}

\author[1]{\fnm{Chuxun} \sur{Liu}}

\author[1]{\fnm{Huangyi} \sur{Kang}}

\author[2]{\fnm{Caihua} \sur{Liu}}

\affil*[1]{\orgdiv{School of Artificial Intelligence}, \orgname{Guilin University of Electronic Technology}, \orgaddress{\state{Guilin}, \country{China}}}

\affil[2]{\orgdiv{Department of Computer Science}, \orgname{Cornell University}, \orgaddress{\state{New York}, \country{United States of America}}}


\abstract{Biomarker discovery from high-throughput transcriptomic data is crucial for advancing precision medicine. However, existing methods often neglect gene–gene regulatory relationships and lack stability across datasets, leading to conflation of spurious correlations with genuine causal effects. To address these issues, a causal graph neural network (Causal-GNN) method that integrates causal inference with multi-layer graph neural networks (GNNs) is developed. The key innovation is the incorporation of causal effect estimation for identifying stable biomarkers, coupled with a GNN-based propensity scoring mechanism that leverages cross-gene regulatory networks. The experimental results demonstrate that our method achieves consistently high predictive accuracy across four distinct datasets and four independent classifiers. Moreover, it enables the identification of more stable biomarkers compared to traditional methods.Our work provides a robust, efficient, and biologically interpretable tool for biomarker discovery, demonstrating strong potential for broad application across medical disciplines.}%, with significant potential for applications in various medical domains.}


\maketitle

\section{Introduction}\label{sec1}
Early diagnosis plays a vital role in clinical medicine by facilitating timely interventions and improving therapeutic outcomes, thereby increasing patient survival rates. Biomarkers ---defined as quantifiable molecular indicators of an organism's physiological or pathological state---serve as fundamental tools in clinical diagnostics~\citep{strimbu2010biomarkers}. Acting as "molecular messengers," they provide a critical bridge between disease mechanisms and clinical applications. Recent research in molecular biology and genomic technologies has further elevated the importance of biomarkers, prompting a transition from traditional morphological assessments toward precision molecular subtyping~\citep{collins2015new}. This transition focuses on molecular biomarkers such as mRNA and miRNA, which identify key genes and signaling pathways involved in disease progression. Such insights support targeted therapy selection, patient stratification, and prognostic evaluation. Numerous studies highlight the translational value of these molecular biomarkers. For example, Jia Liu et al. identified a plasma-based signature of three miRNAs (miR-21, miR-29a, and miR-92a) with high sensitivity and specificity for non-invasive early detection of colorectal cancer~\citep{liu2023three}. Yi Han et al. discovered a plasma extracellular vesicle (EV) mRNA panel that effectively predicts overall postoperative survival in pancreatic ductal adenocarcinoma patients~\citep{han2023plasma}. Kanta Horie et al. showed that circulating levels of MTBRtau243 protein accurately reflect the cerebral accumulation of pathogenic tau aggregates and correlate with the severity of Alzheimer's disease (AD), establishing plasma MTBRtau243 as a key biomarker for quantifying tau pathology in AD~\citep{horie2025plasma}. Collectively, these researches indicate the critical essential role of biomarker identification and characterization in advancing clinical diagnostics, with significant implications for enhancing diagnostic accuracy, guiding treatment strategies, and ultimately improving patient outcomes.

The RNA-seq technology enables the genome-wide quantification of RNA expression levels, providing a powerful platform for the identification of potential biomarkers.Several computational methods have been developed for biomarker identification from RNA-seq data, primarily categorized into three types: wrapper, embedded, and filter methods~\citep{kaur2021feature}. However, recent advancements have increasingly emphasized filter and embedded approaches, reflecting their growing prominence in the field.

Filter methods assess feature importance based on intrinsic characteristics of data, typically offering higher computational efficiency compared to wrapper methods. Representative examples include Multi-Cluster Feature Selection (MCFS)~\citep{cai2010unsupervised}, Orthogonal Least Squares-Based Fast Feature Selection (Fastcan)~\citep{zhang2022orthogonal}, and Deep Feature Screening(DFS)~\citep{li2023deep}. Embedded methods strike a balance between filter and wrapper approaches by integrating feature selection within the model training process, thereby combining advantages from both paradigms. A notable example is Contrastive Feature Selection (CFS-master)~\citep{weinberger2023feature} exemplifies this category, though it demonstrates suboptimal performance in practical feature selection tasks compared to theoretically expected outcomes~\citep{li2017feature, venkatesh2019review, wolf2005feature}. A key limitation of most existing methodologies is that they primarily rely on correlations between disease phenotypes and  molecular features, thereby failing to distinguish genuine causal relationships from confounding associations~\citep{yu2020causality,geer2011correlation}. This limitation reduces biological interpretability, particularly in complex diseases like cancer and neurodegenerative disorders, where non-linear gene interactions and unmeasured confounders complicate  accurate biomarker identification and causal inference~\citep{vanderweele2013definition}. To address this challenge, we previously developed a causal inference method for biomarker discovery in Alzheimer’s disease, aiming to enhance the detection of genuine causal relationships amid intricate biological confounding~\citep{wu2024causal}. However, this approach has several limitations: the propensity scores derived via  multiple regression are inherently constrained to capture only co-regulatory information while neglecting cross-regulatory effects—indirect regulation mediated by intermediate molecules. This limitation diminishes the accuracy of propensity score estimation and restricts the method's ability to encapsulate the full complexity of gene regulatory networks in Alzheimer’s disease.

In this paper, we develop a novel method that integrates Graph Neural Networks (GNNs) with causal inference to analyze high-dimensional biological data. This method consists of three steps: (1) regulatory network construction: a gene regulatory graph is created in which nodes represent genes and edges indicate gene co-expression relationship, with edge weights reflecting regulatory strength; (2) propensity scoring using a three-layer GNN: this model integrates up to three-hop neighborhoods to leverage cross-regulatory signals across modules, generating node-level propensity scores that estimate treatment probabilities based on high-dimensional graph-embedded covariates; (3) estimation of average causal effect : estimating each gene’s average causal effect on the phenotype by  utilizing these propensities. Then ranking all genes through their average causal effects.

Experimental results demonstrate that our method achieves higher accuracy and F1 scores while significantly reducing the feature set. Additionally, the causal estimation step provides more stable and reproducible biomarkers across multiple runs and datasets, highlighting the efficiency, scalability, and suitability of our method for precision medicine applications in biomarker discovery.
%However, using RNA-Seq data for biomarker discovery poses significant challenges due to several intrinsic properties: high dimensionality (with tens of thousands of genes measured per sample), biological and technical heterogeneity (from variations in sample sources, sequencing protocols, and batch effects), and sparsity (many genes exhibit low or no expression in certain samples)~\citep{han2015advanced,zhao2016bioinformatics}. These factors complicate the extraction of meaningful biomarker signals from background noise. To address these challenges, 

Several computational methods have been developed for biomarker identification from RNA-seq data, primarily categorized into three types: wrapper, embedded, and filter methods~\citep{kaur2021feature}. However, recent advancements have increasingly emphasized filter and embedded approaches, reflecting their growing prominence in the field.

\section{Dataset}\label{sec2}
%The data employed in this study mainly stems from the A Blood Platelets-based Gene Expression Database for Disease Investigation (PltDB) and the Gene Expression Omnibus (GEO). %PltDB is a specialized database concentrating on the gene expression profiles of platelets, covering high-throughput gene expression data of various disease types. It provides a plentiful source of information for research related to hematological disorders and other systemic diseases. GEO, on the contrary, is a widely utilized public gene expression database that incorporates transcriptome sequencing and microarray gene expression data from diverse organisms worldwide. 

In this study, we utilized data of breast cancer, non-small cell lung cancer, and glioblastoma from PltDB~\citep{zou2022pltdb}, as well as datasets of AD from GEO (GSE33000 and GSE44770). Detailed information about each dataset can be found in Table~\ref{tab:datasets}.

\begin{table}[!t]
\caption{Dataset Information and Clinical Sample Distribution
\label{tab:datasets}}
\footnotesize
\begin{tabular}{@{}llrrl@{}}
\toprule
Dataset & \multicolumn{1}{c}{Disease} & \multicolumn{2}{c}{Sample Size} & Source \\
\cmidrule(r){3-4}
 & & Diseased & Healthy &  \\
\midrule
Breast Cancer & Breast cancer & 53 & 563 & PltDB \\
NSCLC\textsuperscript{a} & Lung cancer & 453 & 563 & PltDB \\
Glioblastoma & Glioblastoma & 253 & 563 & PltDB \\
GSE33000/44770\textsuperscript{b} & Alzheimer's & 258 & 439 & GEO \\
\bottomrule
\end{tabular}
\footnotetext{Table 1: This table lists the datasets used in the study, including the type of disease, sample size, and the source of the data.}
\footnotetext[a]{Non-Small Cell Lung Carcinoma;}
\footnotetext[b]{Merged GEO datasets;}

%\item PltDB: A Blood Platelets-based Gene Expression Database for Disease Investigation; GEO: Gene Expression Omnibus
\end{table}
%In this study, we utilized data of breast cancer, non-small cell lung cancer, and glioblastoma from PltDB~\cite{}, as well as datasets of Alzheimer’s disease from GEO (GSE33000 and GSE44770). Detailed information about each dataset can be found in Table 2-1. 

We extract only mRNA data from blood RNA-seq samples and discard all other data. Next, we remove low-quality or redundant entries to ensure data reliability and consistency. We then check for missing values and employ mean imputation to reduce their impact on subsequent analyses. To eliminate scale discrepancies across samples, we standardize all gene expression data to a uniform scale.

\section{Methodology}\label{sec3}

%\subsection{A Brief Overview of Causal Inference}\label{subsec3}

%Causal inference aims to identify and quantify the causal effects among variables using observational or experimental data, going beyond simple statistical associations~\citep{glymour2019review,pearl2010causal,pearl2009causal}. 
Causal inference aims to quantify the causal effects among variables using observational data, going beyond simple statistical associations~\citep{glymour2019review,pearl2010causal,pearl2009causal}. 
The core framework for causal discovery in modern causal analysis is the Structural Causal Model (SCM), which is proposed by Judea Pearl~\citep{pearl2010introduction}. This framework contains three stages: (1) constructing a causal graph, (2) controlling for confounding variables, and (3) calculating the average causal effects of each sample. %The overall framework of our method is presented in Fig~\ref{fig3}
%This framework encodes causal hypotheses and reasoning, combines causal graphs with counterfactual logic, and estimates causal effects in three stages: (1) constructing a causal graph, (2) controlling for confounding variables, and (3) calculating the causal effect.

\begin{figure*}[!t]%
\centering
{\includegraphics[width=0.9\linewidth]{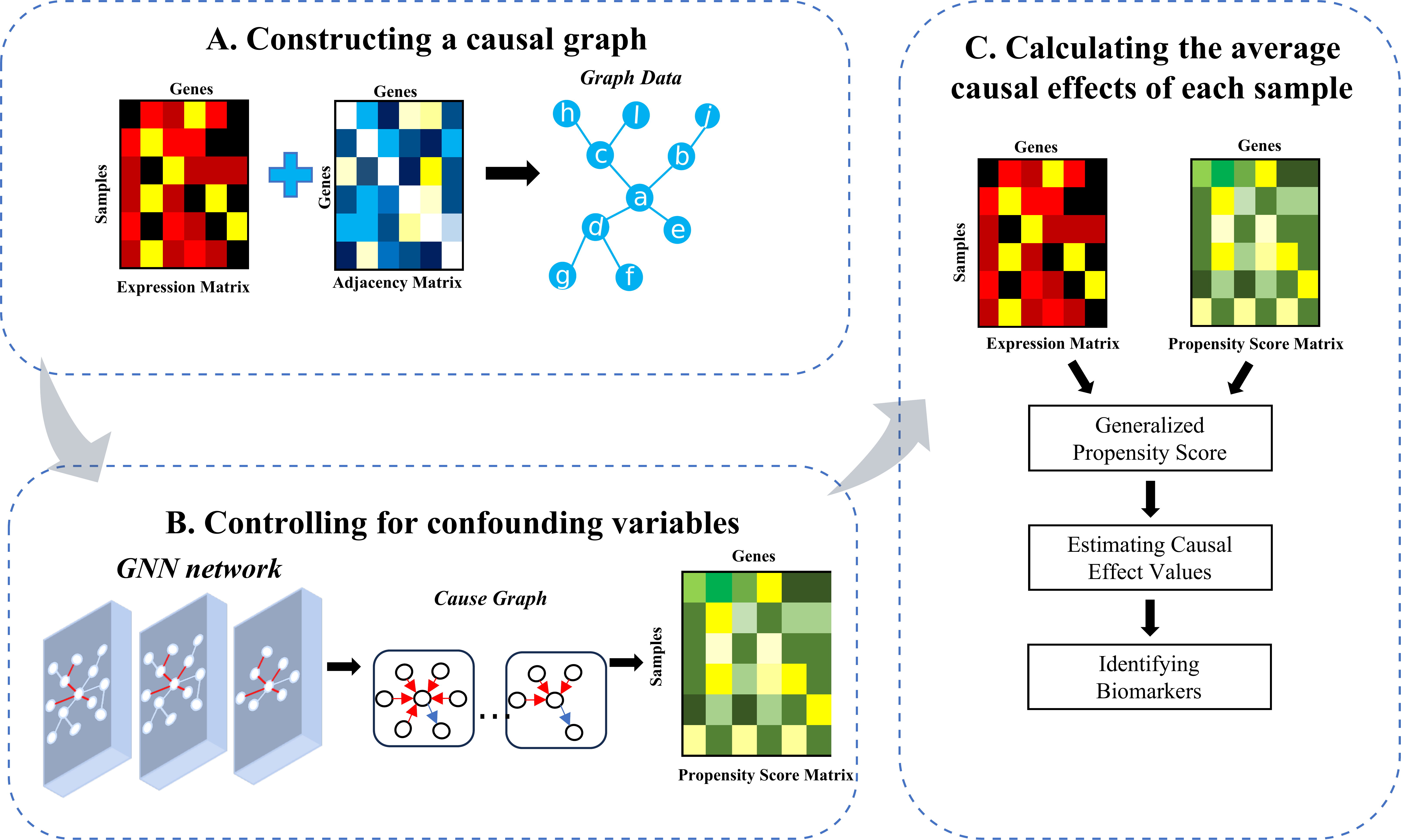}}
\caption{The Framework of Our Methodology. Step1: constructing the gene regulatory network; Step2: calculating propensity score via graph neural network; Step3: calculating the average causal effects of each gene.}\label{fig3}
\end{figure*}

In this paper, we develop the causal graph neural network method to identify disease biomarkers which has the similar framework of causal inference. Our method comprises three key steps: (1) Constructing gene-regulatory network; (2) calculating propensity score via graph neural network; and (3) estimating the average causal effect. The overall framework of our method is presented in Fig~\ref{fig3}.
%\subsubsection{Causal Graph}\label{subsubsec2}
%\subsection{Feature Selection Method for Causal Inference Based on Graph Neural Networks}\label{subsec3}
\subsection{Constructing Gene Regulatory Network}\label{subsec3}
%In this paper, we present a causal inference feature selection method that is enhanced by graph neural network to optimize the construction of causal graphs for the identification of disease biomarkers. The proposed methodology comprises three key steps: (1) data preprocessing; (2) utilizing GNNs for causal graph construction and propensity score calculation; and (3) conducting causal inference to estimate causal effects. The overall framework is illustrated in Fig~\ref{fig3}.

%\textbf{Data Preprocessing:} we extract only mRNA data from blood RNA-seq samples and discard all other data. Next, we remove low-quality or redundant entries to ensure data reliability and consistency. We then check for missing values and employ mean imputation to reduce their impact on subsequent analyses. To eliminate scale discrepancies across samples, we standardize all gene expression data to a uniform scale.

For given a expression profile $\mathbf{X} \in \mathbb{R}^{N \times d}$, where $d$ is the number of samples and $N$ is the number of gene.  We denote that $ G = \{ g_1, g_2, \dots, g_N \}$ is the gene set in $X$ and $Y = \{ Y_1, Y_2, \dots, Y_d \}$ is the label of the sample~(disease or not). $\mathbf{A} \in \mathbb{R}^{N \times N}$ is the adjacency matrix of gene regulatory network and is calculated by the equation~\ref{eq5}:
%if each gene $g_i$ is associated with corresponding expression data, let $\mathbf{X} \in \mathbb{R}^{N \times d}$ denote the expression proflie, where the $i$-th row vector $\mathbf{x}_i$ is the $d$-dimensional feature vector for gene $g_i$.
%Consequently, $\mathbf{X}$ and $\mathbf{A}$ together form the input to the Graph Neural Network
%Let $\{ g_1, g_2, \dots, g_N \}$ denote the set of genes, where $N$ is the total number of genes. 
%The information of gene interaction is obtained from the RNA Inter Database~\cite{}. If two genes $g_i$ and $g_j$ has 
%Gene interaction information, obtained from the RNA Inter Database, can be considered as a collection of edges $\{(g_i, g_j)\}$, indicating interaction or regulatory relationships between two genes. For efficient processing with Graph Neural Networks (GNNs), these relationships are typically represented by an adjacency matrix 
\begin{equation}
A_{ij} = 
\begin{cases}
1, & \text{if } \exists\, \text{interaction between } g_i \text{ and } g_j; \\
0, & \text{otherwise (no interaction).}\label{eq5}
\end{cases}
\end{equation}
The information of gene interaction is obtained from the RNA Inter Database~\citep{kang2022rnainter}. %If gene  $g_i$ is interact $g_j$

\subsection{Calculating Propensity Score via Graph Neural Networks}
%\textbf{Causal graph construction based on graph neural network:}
%if each gene $g_i$ is associated with corresponding expression data, let $\mathbf{X} \in \mathbb{R}^{N \times d}$ denote the feature matrix, where the $i$-th row vector $\mathbf{x}_i$ is the $d$-dimensional feature vector for gene $g_i$. Consequently, $\mathbf{X}$ and $\mathbf{A}$ together form the input to the Graph Neural Network.
In this section, the graph neural network is employed to calculate the propensity score of each mRNA. The propensity score aims to estimate the effect of a mRNA by counting for its co-regulated mRNAs that predict outcomes (disease or not)~\citep{wu2024causal}. Therefore, the propensity score is the probability of the mRNA conditionally on its co-regulated mRNAs. The graph neural network has strong aggregation capabilities for calculating the conditional relationship between mRNA and its co-regulated mRNAs.
%We employ graph neural network in the process of constructing causal graphs. Through graph convolution, the network effectively aggregates information from local neighborhoods and optimizes the structure of the causal graph. Simultaneously, the node representations generated by the network are utilized to compute propensity scores for each gene, resulting in a matrix of gene-level propensity scores.

The graph neural network has many different types of networks, such as graph convolutional network(GCN), graph attention network (GAT), and message passing neural network (MPNN). The core mechanism of the GCN lies in performing spectral-domain convolutions on graph data, integrating adjacency relationships into the feature propagation process. Therefore, the GCN is suitable for estimating the conditional relationship between mRNA and its co-regulated mRNAs. Taking the simplified model by Kipf \& Welling as an example, the propagation in a single-layer GCN in this paper can be represented as equation~\ref{eq2}~\citep{kipf2016semi,defferrard2016convolutional}:
\begin{equation}
\begin{split}
%H^{(l+1)} = \sigma\left( \hat{D}^{-1/2} A \hat{D}^{-1/2} H^{(l)} W^{(l)} 
\text{GCN}_{l+1}(H^{(l)}, A) = \sigma\left( \hat{D}^{-1/2} A \hat{D}^{-1/2} H^{(l)} W^{(l)}
\right)\label{eq2}
\end{split}
\end{equation}
The key parameters in the formula are as follows:
\begin{itemize}
    \item $H^{(l)} \in \mathbb{R}^{N \times d_l}$: Node representations at layer $l$. Where $H^{(0)} = X$.
    \item $D$: The degree matrix of adjacency matrix $A$. Its diagonal transformation $D^{-\frac{1}{2}}$ normalizes the aggregation of neighboring features, mitigating degree bias.
    \item $W^{(l)} \in \mathbb{R}^{d_l \times d_{l+1}}$: The trainable parameter matrix at layer $l$.
    \item $\sigma(\cdot)$: The activation function, here implemented as ReLU.
\end{itemize}
%In the improved GCN proposed in this study, each convolutional layer is immediately followed by batch normalization (BatchNorm) and a nonlinear activation function, with dropout applied selectively to mitigate overfitting. These adjustments significantly enhance the model's robustness and generalization, particularly addressing the high noise and sparsity commonly encountered in gene networks.
The convolutional layer is followed by batch normalization (BatchNorm) and a nonlinear activation function, with dropout applied selectively to mitigate overfitting. These adjustments significantly enhance the model's robustness and generalization, particularly addressing the high noise and sparsity commonly encountered in gene networks.

%To effectively capture nonlinear regulatory patterns from gene expression data and improve the modeling of deep topological structures within gene regulatory networks, a three-layer stacked architecture is designed as follows:
The number of convolutional layer is the farthest distance that node can travel. The capacity for gene regulation diminished with the increasing number of regulatory layers increase. Therefore, a three layer GCN is applied in this paper. The three layer GCN is presented by equation~\ref{eq12}.

\begin{equation}
\begin{split}
H^{(1)} &= \text{GCN}_1(X, A) \\
H^{(2)} &= \text{GCN}_2(H^{(1)}, A) \\
H^{(3)} &= \text{GCN}_3(H^{(2)}, A) + X_{\text{skip}}
\end{split}\label{eq12}
\end{equation}
%The final output of GNN is $O =  X - H^{(3)}$. 
The output of the third layer $H^{(3)}$ is a $N \times d$ matrix and regards as the propensity score matrix. The term $X_{\text{skip}}$ denotes a residual connection that adds the original node features $X$ back to the GCN output, which preserves node-specific information and mitigates over-smoothing. It will be used for estimating the average causal effect.
\subsection{Estimating the Average Causal Effect}
%To improve causal inference accuracy and better identify disease biomarkers, we apply a Pearson correlation coefficient (PCC) filter before conducting the generalized propensity score analysis. For each gene, we extract its copy-number and expression values from the gene expression matrix and the GNN-derived propensity score matrix, respectively, then merge them into a unified data structure. After removing any missing values, we compute the PCC between these two sets of values. Genes with an absolute PCC exceeding a specific threshold (default $\pm 0.8$) are retained for further analysis, as they exhibit strong linear correlation. Specifically, a gene is considered significantly correlated if its PCC is $\ge 0.8$ or $\le -0.8$.
The generalized propensity score is an extension of the propensity score for measuring intervention of continuous values. %Since the expression of gene is  continuous value. %Given the gene set $G$, the generalized propensity score $R_g$ is computed by equation~\ref{eq6}:

%The generalized propensity score is derived from a propensity score matrix $\mathbf{EP}$, an $N \times M$ matrix where $N$ is the number of samples and $M$ is the number of genes. Let $Y = \{y_1, \ldots, y_N\}$ be the sample outcomes (0 for normal, 1 for diseased). 
%For each gene $g$ andits set of  interacting genes $\mathrm{Ne}(g) = \{\mathrm{ne}_1^g, \ldots, \mathrm{ne}_s^g\}$, the generalized propensity score $R_g$ is computed by equation~\ref{eq6}:
%The $g_i = \{ g^i_1, g^i_2, \dots, g^i_d\}$ is the expression level of gene g_i in all samples.
Given a gene $g = \{ g_1, g_2, \dots, g_d\}$,  $H^{(3)}_{g}$ presents the generalized propensity score $R_g$ is computed by equation~\ref{eq6}:
\begin{equation}
%R_g = \tan\bigl(\alpha\,(g - H^{(3)}_{g}) + c\bigr)\label{eq6}
R_g = \tanh\bigl(g - H^{(3)}_{g}\bigr)\label{eq6}
\end{equation}
where $g$ represents the gene expression values, and $f_g$ is the GNN-derived propensity score.

Because the RNA-seq data yield a binary outcome (0 for normal, 1 for diseased),the logistic regression model is employed to estimate each gene's causal effect. The formula of estimating the causal effect of gene $g$ is equation~\ref{eq8}:
\begin{equation}
\text{Logistic}(g) = \frac{1}{1 + e^{-M}}\label{eq7}
\end{equation}
\begin{equation}
M = \bigl(a_0 + a_1 g + a_2 g^2 + a_3 R_g + a_4 R_g^2 + a_5 R_g \cdot g\bigr)\label{eq8}
\end{equation}
Where $a_i$ are parameters of logistic regression model. Once trained, this model clarifies the probability of disease occurrence under the regulation of a given gene. We then define the average causal effect (ACE) for gene $g$ as equation~\ref{eq9}:
\begin{equation}
\text{ACE}(g) = \frac{1}{d} \sum_{i=1}^{d} \bigl(Y_i - \text{Logistic}(g_i)\bigr)^2.\label{eq9}
\end{equation}
%Where $g_i$ is the 
%A lower $\text{ACE}(g)$ indicates a stronger capacity of mRNA $g$ to distinguish between normal and diseased samples. Consequently, all genes are ranked in ascending order by their ACE values.

The lower $\text{ACE}(g)$ of gene $g$, the stronger capacity of this gene in distinguishing normal and diseased samples. Consequently, all genes are ranked in ascending order by their ACE values.

\section{Results and Discussion}\label{sec4}
\subsection{Experimental design and evaluation metrics}\label{subsec4}
%High-quality biomarkers should accurately distinguish between diseased and healthy individuals, resulting in robust classification outcomes. %When evaluating biomarker feature selection methods, the performance of the selected biomarkers in classification tasks is commonly used. 
A small subset of mRNAs could distinguish patients from healthy samples, resulting in robust classification outcomes. These mRNAs can be viewed as biomarkers. The process of identifying biomarkers is that ranking mRNAs by ascending average causal effect, then using a classifier to evaluate their ability to differentiate between patients and healthy samples. 
As the number of selected mRNAs increases, the performance of classifier rises with increasing mRNAs, reaching a maximum when the significant mRNAs are captured. These significant mRNAs are considered as biomarkers. Different classifiers achieve significantly different classification performance, even they use the same features. Therefore, four classifiers—Support Vector Machine (SVM), Naive Bayes (NB), Decision Tree (DT), and XGBoost (XGB)— are selected(The code for this study will be made available at the following repository upon acceptance https://github.com/32713271/Causal-Graph-Neural-Networks-for-Mining-Stable-Disease-Biomarkers). 
%and conducted a comparative analysis against five feature selection and causal inference methods, with CFS-master serving as the baseline~\citep{weinberger2023feature}:
%Considering the potential variations in performance that may arise from different feature selection algorithms and classifiers, we selected four classifiers—Support Vector Machine (SVM), Naïve Bayes (NB), Decision Tree (DT), and XGBoost (XGB)—and conducted a comparative analysis against five feature selection and causal inference methods, with CFS-master serving as the baseline~\citep{weinberger2023feature}:
%gradually improves until it reaches a peak. 
%Finally, the classification performance of the classifier reaches a maximum once all key mRNAs are included. 
%At that point, the identified mRNAs are considered biomarkers. Additionally, the number of selected biomarkers also serves as a performance indicator: fewer identified biomarkers correspond to better classification outcomes.

The core metric of measuring the performance of classifier are F1-score and Precision. A higher F1-score or a higher precision indicates stronger discriminative capability of the selected mRNAs for differentiating between patient and healthy samples~\citep{grandini2020metrics}. The F1 score is calculated by equations \ref{f1}:
\begin{equation}\label{f1}
    F1=2 \times \frac{precision \times recall}{precision+recall}
\end{equation}
Precision: The proportion of true positive samples among all the samples predicted as positive.Recall: The proportion of correctly predicted positive samples among all the actual positive samples.

Additionally, the number of selected biomarkers also serves as a performance indicator. The optimal method maximizes the classification performance while minimizing the cardinality of selected features.

\subsection{Parameter selection}\label{subsec4}
The depth of GNNs refers to the number of successive message-passing within the network. Similarly, the depth of the GCN in our method implies the number of successive cross-regulate genes within the gene regulate network. In order to determine an appropriate depth of GCN our method, we compare three different depth GCN architectures (2-, 3-, and 4-layer). All architectures apply the same input dimensionality derived from the feature matrix and are optimized by the AdamW algorithm with a learning rate of $0.01$ and weight decay of $5e-4$. To ensure training stability, gradient clipping with a maximum norm of 1.0 was applied. Furthermore, a ReduceLROnPlateau scheduler was utilized to dynamically adjust the learning rate in response to stagnation in training loss, with a reduction factor of $0.5$ and patience set to $10$ epochs.

The training process was supervised by a custom-designed ImprovedGraphLoss function, aiming to minimize the discrepancy between the learned node embeddings and the underlying graph topology. The best-performing model during the 500 training epochs was selected based on the minimum training loss.

\begin{figure}[!t]%
\centering
{\includegraphics[width=0.9\linewidth]{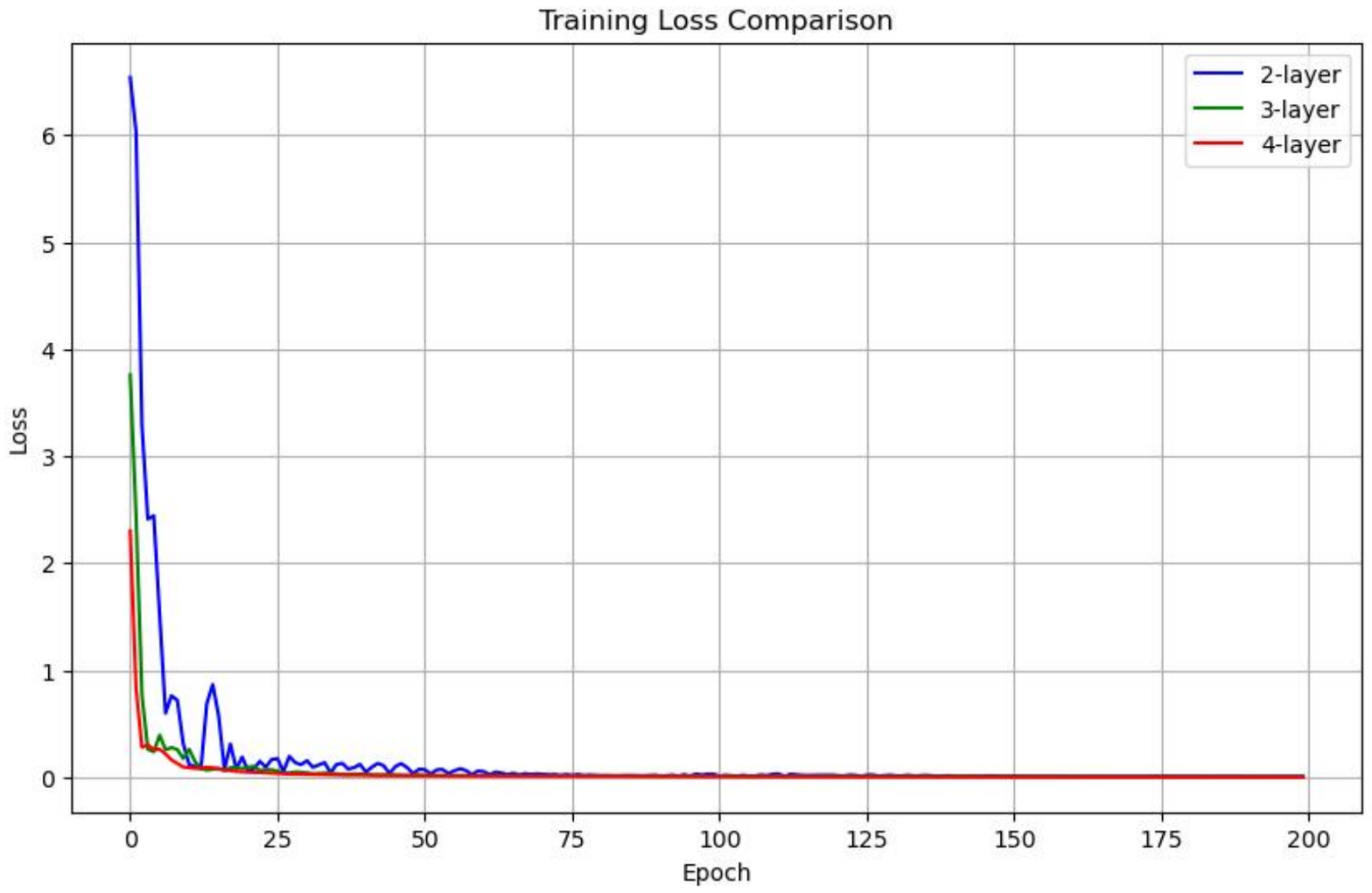}}
\caption{Training loss trajectories of three different GCN architectures with varying network depths (2-layer, 3-layer, and 4-layer) over 200 epochs. }\label{fig4}
\end{figure}

Figure~\ref{fig4} presents the training loss trajectories of these GCN architectures. As illustrated in this figure, the 3-layer GCN architecture consistently outperformed the 2-layer GCN architecture in terms of convergence speed and final loss value, while maintaining better stability than the 4-layer GCN architecture, which showed marginal gains at the cost of increased computational complexity. Consequently, a 3-layer GCN architecture model was adopted in this study as a trade-off between representational power and training efficiency. This experiment implies that as the number of regulatory layers increases, the impact of cross-regulatory genes diminishes. %This choice reflects a balance between model expressiveness and the avoidance of potential overfitting or over-smoothing, which can occur in excessively deep graph neural networks.

%\subsection{Experimental Results and Discussion}\label{subsec4}
\subsection{Comparing the predictive performance with other methods}\label{subsec4}
%According to the experimental design and evaluation metrics outlined previously, this study utilized four RNA-seq datasets (breast cancer, non-small cell lung cancer, glioblastoma, and Alzheimer's disease) for experimentation. The results of these experiments are summarized in Tables 2 through 5, with each table corresponding to a specific dataset.
Many methods for feature selection have been developed. In this paper, we focused on comparing
the performance of our method with five representative feature selection methods in four datasets (breast cancer, non-small cell lung cancer, glioblastoma, and Alzheimer's disease). These five popular methods are :

\begin{enumerate}
\item CFS-master: Feature Selection in the Contrastive Analysis Setting (as the baseline for a state-of-the-art performance benchmark)~\citep{weinberger2023feature}.
\item MCFS: Unsupervised feature selection for multi-cluster data~\citep{cai2010unsupervised}.
\item Fastcan: Orthogonal least squares-based fast feature selection for linear classification~\citep{zhang2022orthogonal}.
\item DFS: Deep feature screening for ultra-high-dimensional data via deep neural networks~\citep{li2023deep}.
\item \textbf{Traditional Causal Inference }:Uses a standard causal inference process (propensity score matching) without incorporating GNN for causal structure learning or parameter optimization~\citep{wu2024causal}.
\end{enumerate}

\textbf{Non-Small Cell Lung Cancer Dataset:} 
In this dataset, the F1-score and precision of our method are higher than other methods across all classifiers,. Although our method do not employ the minimum quantity of feature sets in the Naïve Bayes (NB) and Decision-tree classifiers, predictive performance margins were large. 
%For example, in the decision-tree model, our approach exceeded the minimum quantity baseline, Causal Inference (10 features), by 9.3\% (F1: 0.825 vs. 0.755). Similarly, under the NB classifier, the predictive performance of our method surpassed Causal Inference (23 features) by 10.0\% (F1: 0.867 vs. 0.788). 
The predictive performance of our method surpassed Causal Inference by 9.3\% (F1: 0.825 vs. 0.755) and 10.0\% (F1: 0.867 vs. 0.788) under decision-tree and NB classifiers, respectively.
These results denote that aggressive feature pruning may discard genes essential for accurate classification, whereas our strategy balances parsimony with predictive power.
%In this dataset, our method produced the highest F1-score in every classifier, underscoring its robustness and generalizability. Although it did not employ the most parsimonious feature sets in the naïve Bayes (NB) and decision-tree models, its performance margins were large. In the decision-tree model, for example, our approach exceeded the best feature-parsimonious baseline, Causal Inference (10 features), by 9.3\% (F1: 0.825 vs. 0.755). Likewise, in NB it surpassed Causal Inference (23 features) by 10.0\% (F1: 0.867 vs. 0.788). These results suggest that aggressive feature pruning may discard genes essential for accurate classification, whereas our strategy balances parsimony with predictive power.

\textbf{Alzheimer’s Disease Dataset:} In the Alzheimer’s disease (AD) dataset, the method again excelled under complex class-imbalance conditions. Within the decision-tree model it achieved the highest F1-score (0.948) while selecting only 14 biomarkers—the fewest among all contenders. In NB, its F1-score (0.900) was marginally below that of CFS-Master (0.903), yet it required an order of magnitude fewer features (4 vs. 45), illustrating superior efficiency. Relative to DFS (single-feature), our method improved accuracy by 4.8\% (0.900 vs. 0.859), confirming an advantageous trade-off between performance and parsimony.

\begin{sidewaystable*}[!t]
\centering
\caption{F1-score, accuracy, and number of selected features of six feature-selection methods on four datasets across four classifiers}
\label{tab:results}
\sisetup{table-format=2.0,round-precision=3} % 设置表格数字格式和四舍五入精度
\begin{tabular*}{\textwidth}{@{\extracolsep\fill}ll
S[table-format=2.0] S[table-format=1.3] S[table-format=1.3]
S[table-format=2.0] S[table-format=1.3] S[table-format=1.3]
S[table-format=2.0] S[table-format=1.3] S[table-format=1.3]
S[table-format=2.0] S[table-format=1.3] S[table-format=1.3]}
\toprule
\multirow{2}{*}{Disease} & \multirow{2}{*}{Method}
& \multicolumn{3}{c}{SVM}
& \multicolumn{3}{c}{XGB}
& \multicolumn{3}{c}{NB}
& \multicolumn{3}{c}{DT} \\
\cmidrule(lr){3-5}\cmidrule(lr){6-8}\cmidrule(lr){9-11}\cmidrule(lr){12-14}
& & {Feat.} & {F1} & {Acc} & {Feat.} & {F1} & {Acc}
  & {Feat.} & {F1} & {Acc} & {Feat.} & {F1} & {Acc} \\
\midrule

\multirow{6}{*}{NSCL}
& Our               & {\bfseries 37} & {\bfseries 0.915} & {\bfseries 0.915} & {\bfseries 29} & {\bfseries 0.907} & {\bfseries 0.907} & 47 & {\bfseries 0.867} & {\bfseries 0.868} & 40 & {\bfseries 0.825} & {\bfseries 0.825} \\
& CI  & 46 & 0.856 & 0.856 & 44 & 0.843 & 0.843 & {\bfseries 23} & 0.788 & 0.790 & {\bfseries 10} & 0.755 & 0.755 \\
& CFSM        & 46 & 0.785 & 0.785 & 47 & 0.780 & 0.781 & 50 & 0.710 & 0.711 & 50 & 0.667 & 0.668 \\
& MCFS              & 46 & 0.818 & 0.818 & 47 & 0.799 & 0.799 & 42 & 0.720 & 0.726 & 23 & 0.691 & 0.693 \\
& Fastcan           & 50 & 0.827 & 0.827 & 50 & 0.804 & 0.804 & 50 & 0.711 & 0.713 & 50 & 0.682 & 0.682 \\
& DFS               & 48 & 0.754 & 0.754 & 50 & 0.781 & 0.782 & 47 & 0.688 & 0.690 & 44 & 0.646 & 0.647 \\
\midrule

\multirow{6}{*}{AD}
& Our               & 48 & 0.925 & 0.925 & 34 & 0.962 & 0.963 & 4  & 0.900 & 0.900 & {\bfseries 14} & {\bfseries 0.948} & {\bfseries 0.948} \\
& CI  & {\bfseries 18} & 0.930 & 0.930 & {\bfseries 29} & {\bfseries 0.970} & {\bfseries 0.970} & 10 & 0.881 & 0.879 & 40 & 0.948 & 0.948 \\
& CFSM        & 43 & {\bfseries 0.933} & {\bfseries 0.934} & 32 & 0.963 & 0.963 & 45 & {\bfseries 0.903} & {\bfseries 0.902} & 46 & 0.944 & 0.944 \\
& MCFS              & 40 & 0.860 & 0.864 & 41 & 0.938 & 0.969 & 7  & 0.740 & 0.740 & 50 & 0.901 & 0.901 \\
& Fastcan           & 29 & 0.925 & 0.925 & 35 & 0.967 & 0.967 & 37 & 0.859 & 0.858 & 36 & {\bfseries 0.948} & {\bfseries 0.948} \\
& DFS               & 32 & 0.899 & 0.900 & 41 & 0.960 & 0.960 & {\bfseries 1} & 0.859 & 0.858 & 24 & 0.948 & 0.948 \\
\midrule

\multirow{6}{*}{GBM}
& Our               & {\bfseries 34} & {\bfseries 0.896} & {\bfseries 0.897} & {\bfseries 38} & {\bfseries 0.890} & {\bfseries 0.891} & {\bfseries 7} & {\bfseries 0.863} & {\bfseries 0.865} & {\bfseries 11} & {\bfseries 0.829} & {\bfseries 0.829} \\
& CI  & 49 & 0.850 & 0.852 & 41 & 0.860 & 0.862 & 29 & 0.795 & 0.800 & 47 & 0.793 & 0.794 \\
& CFSM        & 44 & 0.841 & 0.843 & 47 & 0.848 & 0.849 & 40 & 0.801 & 0.802 & 46 & 0.768 & 0.768 \\
& MCFS              & 37 & 0.827 & 0.829 & 39 & 0.811 & 0.814 & 42 & 0.806 & 0.811 & 39 & 0.762 & 0.765 \\
& Fastcan           & 38 & 0.877 & 0.878 & 39 & 0.885 & 0.886 & 43 & 0.830 & 0.831 & 41 & 0.788 & 0.788 \\
& DFS               & 48 & 0.846 & 0.848 & 48 & 0.838 & 0.840 & 15 & 0.769 & 0.769 & 48 & 0.733 & 0.734 \\
\midrule

\multirow{6}{*}{BC}
& Our               & 43 & {\bfseries 0.974} & {\bfseries 0.976} & 48 & {\bfseries 0.975} & {\bfseries 0.976} & 30 & {\bfseries 0.978} & {\bfseries 0.978} & {\bfseries 8} & {\bfseries 0.955} & {\bfseries 0.956} \\
& CI  & {\bfseries 13} & 0.969 & 0.971 & 46 & 0.973 & 0.973 & {\bfseries 14} & 0.972 & 0.971 & 16 & 0.950 & 0.951 \\
& CFSM        & 45 & 0.953 & 0.958 & 45 & 0.962 & 0.967 & 46 & 0.949 & 0.947 & 45 & 0.928 & 0.931 \\
& MCFS              & 36 & 0.949 & 0.953 & 44 & 0.963 & 0.964 & 38 & 0.958 & 0.958 & 35 & 0.920 & 0.920 \\
& Fastcan           & 46 & 0.969 & 0.971 & 44 & 0.973 & 0.973 & 49 & 0.967 & 0.967 & 11 & 0.952 & 0.951 \\
& DFS               & 31 & 0.961 & 0.964 & {\bfseries 35} & 0.968 & 0.969 & 28 & 0.958 & 0.956 & 41 & 0.945 & 0.947 \\
\bottomrule
\end{tabular*}
\footnotetext{Abbreviations: SVM, support vector machine; NB, Naive Bayes. “Feat.” denotes the number of selected features. DT,Decision Tree. CFSM, CFS-Master. Acc, Accuracy. F1, F1-score.  Bold numbers indicate the best result within each dataset and classifier.}
\end{sidewaystable*}

\textbf{Glioblastoma Dataset:}Our algorithm simultaneously maximized accuracy and minimized feature count across all four classifiers. The decision-tree model showed the largest gain: accuracy improved by 4.5\% over Causal Inference (0.829 vs. 0.793) while reducing the feature count by 77\% (11 vs. 47). In NB, accuracy was 4.0 \% higher than Fastcan (0.863 vs. 0.830) with an 84\% decrease in features (7 vs. 43).

\textbf{Breast Cancer Dataset:}On this dataset, our method achieved the highest classification performance (F1-score) in all four models. In the decision-tree classifier, the biomarker panel selected by our approach—only eight features—yielded the best results while keeping the feature count to a minimum. In the SVM, XGBoost, and Naïve Bayes (NB) classifiers, we retained slightly more biomarkers than some competing methods to maximize predictive power. For instance, under NB classifier, our method employed 30 features—16 more than Causal Inference (14)—and these additional variables raised the F1-score from 0.972 to a dataset-leading 0.978. These results show that the proposed approach consistently attains the optimal F1-score, either with a highly compact feature set or with only modest, strategically chosen expansions when necessary.

Across four heterogeneous transcriptomic cohorts, our method consistently delivered very high predictive accuracy under four classifiers and maintained compact biomarker sets. Its gains were most pronounced in the NSCLC and GBM datasets, where it outperformed all baselines both in accuracy and parsimony. In the BC dataset, it achieved a joint optimum of performance and efficiency in the decision-tree model and led accuracy tables compare with other models. In the AD dataset, it provided the best decision-tree performance and the most favorable accuracy–feature trade-off in the remaining classifiers. Collectively, these findings establish the method as a reliable and efficient tool for biomarker discovery in heterogeneous genomic data.

\subsection{Comparing the stability of biomarkers with other methods}\label{subsec4}

The stability of feature selection pertains to the insensitivity of feature selection algorithms to variations in the dataset~\citep{moradmand2025graph,huang2021feature}. This implies that highly similar subsets of features can still be identified across different resampling or batches. In bioinformatics, the high-dimensional data is often accompanied by a low sample size, thereby the stability of biomarkers becomes particularly critical~\citep{moradmand2025graph}. Traditional feature selection methods that rely on correlation tend to produce markedly different candidate gene sets when applied to varying data batches or during resampling processes. Such instability directly undermines the reproducibility of research findings, diminishes their biological validity, and may even result in erroneous conclusions regarding key biomarkers~\citep{moradmand2025graph}.A robust method ensures that the identified features are not merely artifacts of random sampling but possess genuine biological relevance, thereby enhancing the credibility of subsequent biomarker validation and clinical translation.

To quantify the stability of the our method, we employed a robustness assessment protocol in non-small cell lung carcinoma (NSCLC) data. This dataset was randomly partitioned into five disjoint subsets, and feature selection was performed independently on each subset. Stability was evaluated by analyzing the absolute count of overlapping features in the top-50 biomarkers across different group combinations. For each possible combination of group size (where group size ranges from 2 to 5), we computed the overlap count defined as:

\begin{equation}
\text{OverlapCount}(\Omega_g) = \left| \bigcap_{i=1}^{g} S_i \right|
\label{eq:count}
\end{equation}

where $g$ denotes the group size, $S_i$ represents the set of top-50 features from partition $i$, and $\bigcap$ denotes set intersection. 

As shown in Table~\ref{tab:stability}, our method demonstrated superior stability across all group complexity levels. In pairwise comparisons (10 combinations), our method maintained overlap counts between 17--30 features, with an average of 23.2 features per group combination. This performance has competitive with MCFS method, which averaged 25.7 features across the same pairs. 
%with both methods outperforming traditional causal approaches. 
CFS-master never exceeded 2 overlapping features, while DFS remained below $5$ in almost all comparisons (peaking at $9$ in one pairwise case). Our previous causal inference method is not mentioned since the number of the sample in each sub-dataset is too small to calculate the causal inference of each features.

As the number of group combinations increases, the stability performance of our method becomes better than other methods. In the triple-group combinations (10 combinations), our method maintained 12--21 overlapping features, yielding an average of 15.1 features -- 9.4\% higher than MCFS's average of 13.8 features. In quadruple-group combination comparisons (5 combinations), our method achieved 9--15 overlapping featureswith an average of 11.2 features , outperforming MCFS by 75.0\% relative improvement. Most importantly, at the maximum complexity of all five groups, our method preserved 9 common features -- nine times more than MCFS (1 feature). %Othemethods with zero overlap
%The stability advantage became progressively more pronounced in larger group combinations:
%\begin{itemize}
%    \item For triple-group combinations (10 combinations), our method maintained 12--21 overlapping features (24.0\%--42.0\%), yielding an average of 15.1 features (30.2\%) -- 9.4\% higher than MCFS's average of 13.8 features (27.6\%)
%    \item In quadruple-group comparisons (5 combinations), our method achieved 9--15 overlapping features (18.0\%--30.0\%) with an average of 11.2 features (22.4\%), outperforming MCFS by 75.0\% relative improvement
%    \item Crucially, at the maximum complexity of all five groups, our method preserved 9 common features (18.0\%) -- nine times more than MCFS (1 feature) and infinitely more than methods with zero overlap
%\end{itemize}

These results indicate that as increasing the number of group combinations reduces the ability to identify common biomarkers. Compared to other methods, our approach is less affected by the number of group combinations and consistently identify more stable biomarkers. Furthermore, it provides computational stability not achieved by traditional approaches, which often fail to yield reliable biomarker signatures.

%consistent monotonic stability preservation as group combination increases; 2) Progressive performance differentiation from alternatives, with relative advantage increasing from near-parity in pairwise to 900\% superiority in full-group comparison; 3) Reliable computational stability absent in traditional methods, which failed catastrophically (CFS-master: $24/26$ combinations yielded 0 overlap (only two pairwise comparisons had $ \leq 2$ overlaps); DFS: maximum $9$ features only in  pairwise comparison).

%The results demonstrate three fundamental advantages: 1) Consistent monotonic stability preservation as group complexity increases (60.0\% max pairwise $\rightarrow$ 42.0\% triple $\rightarrow$ 30.0\% quadruple $\rightarrow$ 18.0\% full group); 2) Progressive performance differentiation from alternatives, with relative advantage increasing from near-parity in pairwise to 900\% superiority in full-group comparison; 3) Reliable computational stability absent in traditional methods, which failed catastrophically (CFS-master: $24/26$ combinations yielded 0 overlap (only two pairwise comparisons had $ \leq 2$ overlaps); DFS: maximum $9$ features in one pairwise comparison).

\begin{table*}[!t]
\caption{Stability Analysis of Feature Selection Methods Across Group Combinations}
\label{tab:stability}
\centering
\renewcommand{\arraystretch}{1.2}
\begin{tabularx}{\textwidth}{>{\centering\arraybackslash}m{2.2cm} >{\centering\arraybackslash}m{2.5cm}
>{\centering\arraybackslash}X >{\centering\arraybackslash}X 
>{\centering\arraybackslash}X >{\centering\arraybackslash}X
>{\centering\arraybackslash}X}
%\begin{tabularx}{\textwidth}{>{\centering\arraybackslash}m{0.7cm} >{\centering\arraybackslash}X 
%>{\centering\arraybackslash}m{0.5cm} >{\centering\arraybackslash}m{0.7cm} 
%>{\centering\arraybackslash}m{1.0cm} >{\centering\arraybackslash}m{1.0cm} 
%>{\centering\arraybackslash}m{0.5cm}}
\toprule
\textbf{Analysis Type} & \textbf{Group Combination} & \textbf{Our} & \textbf{MCFS} & \textbf{Fastcan} & \textbf{CFS-master} & \textbf{DFS} \\
\midrule

\multirow{10}{*}{\textbf{Pairwise}} 
& Group 1,2 & 25 & 24 & 4 & 0 & 1 \\
& Group 1,3 & 26 & 30 & 0 & 0 & 4 \\
& Group 1,4 & 17 & 22 & 3 & 0 & 0 \\
& Group 1,5 & 30 & 23 & 4 & 0 & 9 \\
& Group 2,3 & 21 & 28 & 2 & 1 & 2 \\
& Group 2,4 & 20 & 22 & 5 & 0 & 1 \\
& Group 2,5 & 21 & 25 & 4 & 2 & 3 \\
& Group 3,4 & 22 & 30 & 1 & 0 & 1 \\
& Group 3,5 & 27 & 29 & 2 & 0 & 5 \\
& Group 4,5 & 23 & 24 & 1 & 0 & 1 \\
& \textbf{Average } & 23.2 & \textbf{25.7} & 2.6 & 0.3 & 2.7 \\
\midrule

\multirow{10}{*}{\textbf{Triple}} 
& Group 1,2,3 & 17 & 17 & 0 & 0 & 0 \\
& Group 1,2,4 & 12 & 11 & 1 & 0 & 0 \\
& Group 1,2,5 & 17 & 11 & 1 & 0 & 1 \\
& Group 1,3,4 & 13 & 19 & 0 & 0 & 0 \\
& Group 1,3,5 & 21 & 14 & 0 & 0 & 3 \\
& Group 1,4,5 & 14 & 9  & 1 & 0 & 0 \\
& Group 2,3,4 & 15 & 15 & 1 & 0 & 0 \\
& Group 2,3,5 & 16 & 16 & 0 & 0 & 0 \\
& Group 2,4,5 & 12 & 10 & 1 & 0 & 0 \\
& Group 3,4,5 & 14 & 16 & 0 & 0 & 0 \\
& \textbf{Average } & \textbf{15.1} & 13.8 & 0.5 & 0 & 0.4 \\
\midrule

\multirow{5}{*}{\textbf{Quadruple}} 
& Group 1,2,3,4 & 11 & 9  & 0 & 0 & 0 \\
& Group 1,2,3,5 & 15 & 6  & 0 & 0 & 0 \\ 
& Group 1,2,4,5 & 9 & 3  & 1 & 0 & 0 \\
& Group 1,3,4,5 & 11 & 7  & 0 & 0 & 0 \\
& Group 2,3,4,5 & 10 & 7  & 0 & 0 & 0 \\
& \textbf{Average } & \textbf{11.2} & 6.4 & 0 & 0 & 0 \\
\midrule

\textbf{All Groups} & All 5 Groups & 9 & 1 & 0 & 0 & 0 \\
\bottomrule
\end{tabularx}

\vspace{1mm}
{\raggedright \footnotesize
The percentage calculation is based on the total number of top 50 features, and the cross-validation adopts a 5-group strategy.Symbol description: Pairwise (two-group), Triple (three-group), Quadruple (four-group).
\par}
\end{table*}

\subsection{The Relationship between Biomarkers and Diseases}\label{subsec4}

To validate the effectiveness of the selected biomarkers, this study conducted Gene Ontology (GO) biological process enrichment analysis on the top 100 mRNAs from Glioblastoma datasets. We identified several significantly enriched gene clusters (FDR $<$ 0.05) associated with key pathological mechanisms of Glioblastoma cancers. The detailed analysis is as follows:

Ribosome Biogenesis and Tumor Proliferation: Genes encoding ribosomal proteins, such as \textit{RPS15A}, \textit{RPS27}, and \textit{RPL21}, are significantly enriched in "Cytoplasmic Translation" (FDR = $3.09 \times 10^{-11}$) and "Ribosomal Small Subunit Biogenesis" (FDR = $5.21 \times 10^{-2}$). Aberrant activation of ribosome biogenesis is a prominent feature of glioblastoma, where it enhances tumor cell proliferation by upregulating nucleolar rDNA transcription ~\citep{tao2023novel}. Studies have shown that ribosomal protein genes are abnormally expressed in glioblastoma cells.

Platelet Aggregation and Tumor Invasive Microenvironment: Genes such as \textit{ACTN1}, \textit{MYH9}, and \textit{GP1BA} are significantly enriched in "Platelet Aggregation" (FDR = $3.24 \times 10^{-6}$). Platelets can influence the tumor microenvironment through interactions between platelet-associated factors and tumor cells ~\citep{xue2024interaction}. Glioblastoma stem cells promote microtubule formation in glioblastoma by transforming growth factor-$\beta$ (TGF-$\beta$), enhancing tumor cell invasion ~\citep{miletic2021tgf}.

Cytoskeletal dynamics and invasion phenotype: Genes like \textit{TPM1}, \textit{MYH9}, and \textit{ACTN1} are enriched in "Actin Filament Organization" (FDR = $8.98 \times 10^{-3}$). Research indicates that Myosin IIA and IIB are the most common isoforms of Myosin II in glioblastoma, associated with glioblastoma cell malignancy ~\citep{picariello2019myosin}.

\section{Conclusion}\label{sec5}
This study introduces a GNN-augmented causal-effect framework that simultaneously learns gene-network structures and quantifies each gene’s causal impact, facilitating principled feature pruning. A key advantage of this framework is its use of a multi-layer graph neural network, which effectively captures richer cross-regulatory information—particularly indirect regulatory effects through intermediate molecules. This capability significantly enhances the accuracy of propensity score fitting, addressing the limitations of traditional methods that only consider co-regulatory information.

Moreover, integrating causal inference methodologies improves the stability of identified biomarkers by distinguishing genuine causal relationships from spurious associations and accounting for confounding factors. This ensures that selected biomarkers are not artifacts of data variability but possess robust biological relevance. Rigorous stability analysis using five-fold resampling on the NSCLC dataset confirms this robustness, with a mean pairwise overlap of $46.4\% \pm 8.2\%$ among selected gene sets. Thus, the framework provides compact, reproducible, and causally interpretable biomarker panels, serving as a powerful tool to advance precision medicine discovery.

\section*{Declarations}

\begin{itemize}
    \item \textbf{Funding:} No funding was received for this work.
    \item \textbf{Conflict of interest/Competing interests:} The authors declare no competing interests.
    \item \textbf{Ethics approval and consent to participate:} Not applicable. This study did not involve human participants, and the data used was publicly available.
    \item \textbf{Consent for publication:} Not applicable. The manuscript does not contain any individual data that requires consent for publication.
    \item \textbf{Data availability:} The dataset(s) supporting the conclusions of this article is(are) available in the PltDB repository (https://www.pltdb-hust.com) and the GEO database (https://www.ncbi.nlm.nih.gov/geo/) (GSE33000 and GSE44770).
    \item \textbf{Materials availability:} Not applicable. The materials used in the study are publicly available datasets.
\end{itemize}


\end{document}